\documentclass[sigconf, authorversion]{acmart}

\AtBeginDocument{%
  \providecommand\BibTeX{{%
    \normalfont B\kern-0.5em{\scshape i\kern-0.25em b}\kern-0.8em\TeX}}}

\copyrightyear{2021}
\acmYear{2021}

\usepackage{float}
\usepackage{natbib}
\usepackage{graphicx}
\usepackage{comment}
\usepackage[utf8]{inputenc}

\begin{document}
\UseRawInputEncoding 
\title{Understanding parents' perceptions of children's cybersecurity awareness in Norway}

\author{Farzana Quayyum}
\authornote{Corresponding author}
\email{farzana.quayyum@ntnu.no}

\author{Jonas Bueie}
\email{jonasbue@stud.ntnu.no}

\author{Daniela S. Cruzes}
\email{daniela.s.cruzes@ntnu.no}

\author{Letizia Jaccheri}
\email{letizia.jaccheri@ntnu.no}

\author{Juan Carlos Torrado Vidal}
\email{Juancarlos.Torrado@uib.no}
\affiliation{%
  \institution{Norwegian University of Science and Technology (NTNU)}
  \city{Trondheim}
  \country{Norway}
}


\renewcommand{\shortauthors}{F. Quayyum et al.}


\begin{abstract}
Children are increasingly using the internet nowadays. While internet use exposes children to various privacy and security risks, few studies have examined how parents perceive and address their children's cybersecurity risks. To address this gap, we conducted a qualitative study with 25 parents living in Norway with children aged between 10 to 15. We conducted semi-structured interviews with the parents and performed a thematic analysis of the interview data. The results of this paper include a list of cybersecurity awareness needs for children from a parental perspective, a list of learning resources for children, and a list of challenges for parents to ensure cybersecurity at home. Our results are useful for developers and educators in developing cybersecurity solutions for children. Future research should focus on defining cybersecurity theories and practices that contribute to children's and parents' awareness about cybersecurity risks, needs, and solutions.
\end{abstract}



\maketitle

\section{Introduction}
Children are becoming increasingly global citizens who use the internet daily and frequently for education, entertainment, and communication. While online, children face a significant number of opportunities that unlock the potential of individuals, technology, and collaboration to create a positive societal impact\footnote{What is Social Good?. https://www.socialchangecentral.com/what-is-social-good}. While online, children face risks as well. Thus, children must learn to use computers and the internet in a safe and secure manner\footnote{Informatics and Digital Skills. https://www.informaticsforall.org/informatics-digital-skills/}. However, children are often unaware of the cybersecurity risks they may be exposed to when using the internet. The International Telecommunication Union (ITU) defines cybersecurity as "the collection of tools, policies, security concepts, security safeguards, guidelines, risk management approaches, actions, training, best practices, assurance, and technologies that can be used to protect the cyber environment and organization and user's assets\footnote{https://www.itu.int/en/ITU-T/studygroups/com17/Pages/cybersecurity.aspx}". Cybersecurity awareness is defined as "a methodology to educate internet users to be sensitive to the various cyber threats, and the vulnerability of computers and data to these threats \cite{rahim}". Without cybersecurity awareness, children are less likely to understand the risks and the implications of their actions on their or others' lives \cite{p84}. Thus, adults especially parents have a responsibility to help children to use the internet safely and make them aware of cybersecurity risks and consequences.

Parents recognize the importance of their role as the mediators of their children's technology use and online security \cite{p98}. Parents control the resources available to their children and manage the environment to protect them from harmful social influences, including the potentially negative effects of adolescent internet use \cite{sonia, p65}. Considering the significant role parents can play in their children's cybersecurity awareness, it is essential to determine what parents think about their children's cybersecurity. Researchers have studied different parenting styles and the role of parents in shaping children's online behavior and safety \cite{VALCKE2010454, rode}. However, little discussion has focused on parents' perceptions of their children's cybersecurity awareness needs. Managing technology use in the home is a dynamic endeavor that has changed over time, presenting new concerns and challenges for parents. Moreover, demographic changes, culture, and socioeconomic status may also impact the socially structured patterns of parenting in the face of children's cybersecurity needs \cite{elstad2014social}.  

In our study, we recognize parents as important stakeholders in raising cybersecurity awareness among children. We investigated parents' cybersecurity concerns, the cybersecurity awareness needs of children, and the cybersecurity challenges for parents in Norway. We addressed the following research questions. \\
RQ1. What are the cybersecurity awareness needs of children?  \\
RQ2. How much do children currently know about cybersecurity, and from which sources have they gained that knowledge?  \\
RQ3. What challenges do parents face in ensuring cybersecurity at home? \\

To answer our research questions, we conducted semi-structured interviews with parents. We presented the findings in terms of children's awareness needs, learning sources, and challenges for parents.

The structure of the paper is as follows: Section \ref{background} introduces related studies about children's cybersecurity awareness, with a focus on the role of parents. Section \ref{researchmethod} describes this study's research method, data collection, and analysis, as well as ethical issues. Results are presented in Section \ref{results}, organized around the three research questions. Section \ref{discussion} presents the finding of this study, and Section \ref{conclusion} concludes the paper.

\section{Background} \label{background}
Awareness of cybersecurity is vital for children to safely use the internet and emerging technologies. In recent years, children's cybersecurity has gained much attention, both from academia and from the tech industry. Researchers have proposed and designed various training programs and games for children to raise their awareness of cybersecurity. For example, \cite{p21} has developed a game for primary and secondary-school children to enhance awareness of privacy on social networks. \cite{p6} has developed an ontology of age-based "best practice" password principles for children. With this ontology, educators and parents can instruct children about password-related principles. Many other studies have also worked to raise cybersecurity awareness among children, including \cite{p14, p48, p24}. 

Many researchers have explored other topics related to cybersecurity for children. For example, \cite{p98} has explored the perspectives of different stakeholders in educating children about internet safety. Studies have explored the role of parents in ensuring their children's security online \cite{p49, p65, p73}. Other studies have investigated how families manage cybersecurity in the home \cite{p5} and family preferences concerning children's online privacy and data mining \cite{p84}. Some researchers have also investigated the strategies parents use to mitigate their children's cybersecurity risks \cite{p5, p55, rode}. Overall, it is clear that parents can play a significant role in their children's online behavior and cybersecurity awareness. However, most of these prior studies have not discussed parental views and preferences when developing children's cybersecurity awareness solutions. When developing any solution for children, we believe it is important to determine what parents think and prefer regarding their children's cybersecurity awareness.

Studies have shown that different parenting styles can influence children's behavior \cite{p65, p73, clausen1996parenting}. Also, the effects of different parenting styles on children have been shown to vary within and across cultures \cite{marc}. In addition, parenting styles differ between countries, and studies have shown that Scandinavian parents are usually less authoritarian towards their children \cite{marc}. Our study extends related work by examining parents' perceptions of their children's cybersecurity awareness in a Scandinavian country (i.e., Norway). We found no existing study that investigated parents' perspectives about their children's cybersecurity in Norway. Therefore, we build on current research by exploring children's cybersecurity awareness needs from their parents' perspectives. We also explore how children learn about cybersecurity and whether parents face any challenges in ensuring cybersecurity at home. 

\section{Methods} \label{researchmethod}
We conducted semi-structured interviews with 25 parents living in Norway who have children aged between 10 and 15 years. We used semi-structured interviews to allow our interviewees to respond in detail to our questions and to introduce issues of their own that they consider relevant to our research. We performed a thematic analysis of the interview data following the steps recommended by \cite{cruzes-dyba}. We applied an integrative approach to the data analysis. We approached the data with specific questions (the research questions) in mind that we wished to code according to. We also referred to the interviewees' concepts. The coding process resulted in 75 codes with 202 references. Afterward, the codes were categorized into themes based on our research questions, keeping the interviewees' concepts unchanged. For example, in Figure \ref{fig:knowledge}, "cybersecurity awareness needs for children" is a higher-order theme that comprises seven themes: technical knowledge, privacy, online content, online contacts, basic knowledge of cybersecurity, positive skepticism, and cyberbullying and bad online attitude. These themes emerged from the associated codes. For example, the "technical knowledge" theme emerged from codes such as information flow in the system, how the system works, viruses and malware, and the possibility of hacking. Similarly, in Figure \ref{fig:challenges}, "challenges for parents" is a higher-order theme that comprises five themes and several codes.  

To develop our interview questions, we used the Goal Question Metric (GQM) approach proposed by \cite{basili}. Our goal was to gain a clear understanding of parents' perceptions of their children's cybersecurity. Thus, our interview questions emphasized parents' opinions and concerns about their children's cybersecurity, their own experiences at home, and the underlying processes of ensuring children's cybersecurity, such as understanding the strategies they use at home and related challenges. The interviews were conducted between July and August 2020. All the interviews were audio-recorded, and the interviewees were asked to give explicit consent in advance of the interviews. Consent was given most often in writing, but, in some cases, consent was also given orally.

To find the interview subjects, we adapted the self-selection and snowball strategy. We contacted people acquainted with our research team members. These contacts had children in the relevant age group. We described our study to them and asked if they wished to be interviewed. To achieve a high degree of diversity, we did not limit our sample to any specific gender, profession, or educational background. We interviewed those who agreed to participate in this study. Some of the interviewees also suggested other people who they thought might be interested in participating. We contacted the potential interviewees by email or SMS and invited them to interview. In all, we interviewed 25 parents. Three interviews were conducted in person, and the rest were conducted either online (using Microsoft Teams, Zoom, or FaceTime) or over the phone. The interviews varied in duration from 10 minutes to 34 minutes. The participants included eight fathers and 17 mothers, and their ages ranged from 38 to 56 years. They had a collective 26 children, aged from 10 to 15 years (six girls and 20 boys). 

Before beginning data collection, we applied for approval from the Norwegian Centre for Research Data (NSD)\footnote{https://www.nsd.no/en/about-nsd-norwegian-centre-for-research-data/}. Moreover, during the interviews, the participants were instructed to avoid sharing personal information about any third party. 

\section{Results} \label{results}
This section presents the results of this study, organized around the research questions. 

\subsection{RQ1. What are the cybersecurity awareness needs of children?} 
\begin{figure}[h!]
\includegraphics[scale=1]{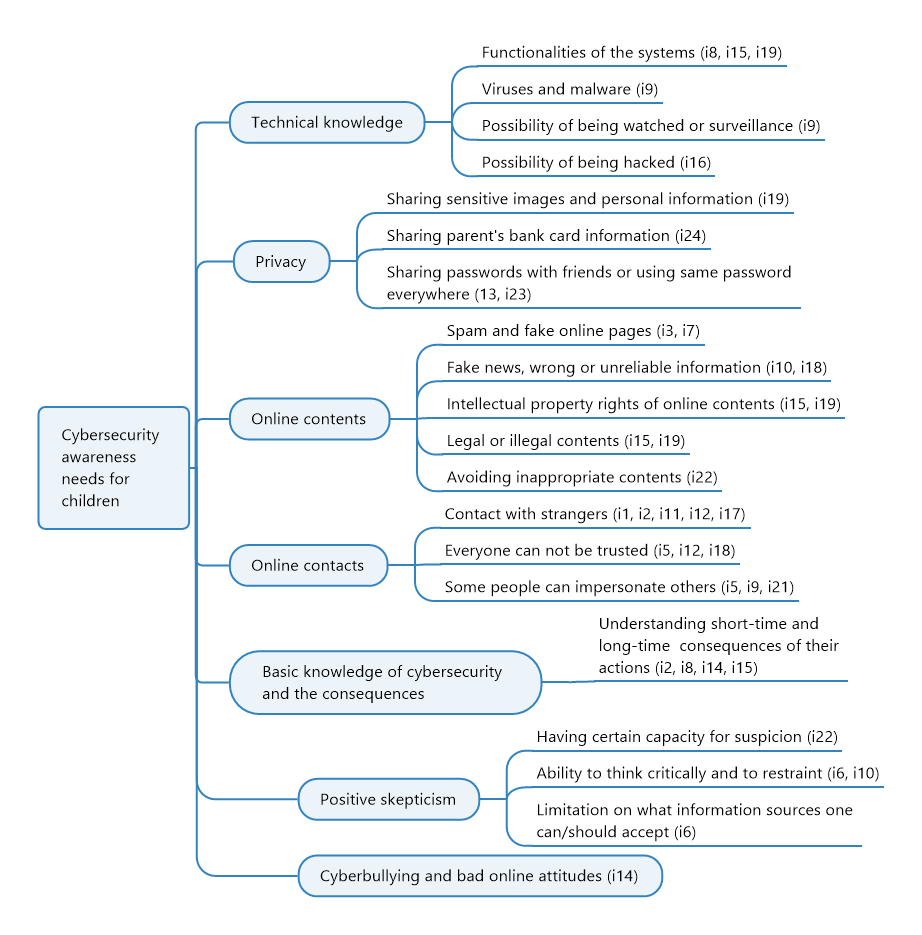}
\caption{Cybersecurity awareness needs for children (from parent's perspectives)}
\Description{A list of cybersecurity awareness needs for children}
\label{fig:knowledge}
\end{figure}

We asked parents to share their cybersecurity concerns regarding their children's use of the internet. We also asked them to share their opinions about the kind of awareness that children need to stay safe online. Thus, we identified some needs that parents believed should be met to ensure their children's online safety, which we present in Figure \ref{fig:knowledge}. 

The parents expressed that their children need both awareness and behavioral attributes to stay safe online. They believed that children should know and understand the technology they use; for example, how the system works, how the information disseminates through the system, and what information is stored. Based on the parents' responses, children also need to know about computer viruses and malware; the quality, reliability, and legality of the material they see and use on the internet; and the identity of their online contacts. A good understanding of their privacy is also important. When we asked the parents about what kind of awareness they thought children need when they use the internet, one parent stated,

{\itshape "It's both fundamental to understand how things are stored and how systems work. [...] It also concerns laws and regulations, i.e. what is actually legal and illegal." (i15)} 

Along with awareness, parents believed children should also have some behavioral attributes. For example, they should be somewhat skeptical about any information they find on the internet, think critically, and refrain from action if something does not feel right. Children also need to be aware of their attitudes when they post or comment on anything online. They should also avoid behaviors that can lead to cyberbullying, and they need to consider the consequences of their actions before they act. One parent described these needs as follows:

{\itshape "[Children should have a] Critical sense of what is there, and you should not press 'yes' and 'ok' on everything." (i10)}

\subsection{RQ2. How much do children currently know about cybersecurity, and from which sources have they gained that knowledge?} 
The parents of younger children reported that their children had some basic awareness about cybersecurity. However, the parents of adolescents reported that their children knew quite a bit about cybersecurity and its risks. Children learned about cybersecurity both from home and school. Children were trained at school on security topics such as sharing personal information, safe password practices, how to handle cyberbullying, sexual abuse, and social media. 

\subsubsection{Discussing cybersecurity at home} The majority of the parents (15 of the 25) explicitly mentioned that they discuss online security issues with their children at home. Their discussions arise from their own experiences or from cybersecurity cases in the media. They try to help their children understand what happened to others and what could happen to them, as well as what they could do if something happened to them. Thus, parents aim to strengthen their children's awareness of the consequences of cybersecurity risks. Some parents mentioned that they also show interest and engage in their children's online activities. This way, they can both monitor their children's activity and discuss cybersecurity issues. 

\subsubsection{Discussing cybersecurity at school}
Norwegian schools play an important role in fostering cybersecurity awareness among children. Cybersecurity is not formally included in Norway's curricula, but schools conduct awareness trainings on various security topics. To prepare and conduct these programs, schools coordinate with other organizations that work to raise cybersecurity awareness among children. From the interviews, we learned about three such organizations that schools work with to train children about cybersecurity and online etiquette: Barnevakten\footnote{https://www.barnevakten.no/}, Bruk Hue\footnote{https://brukhue.no/}, and Medietilsynet\footnote{https://medietilsynet.no}. These organizations help schools by providing training and lectures to students and teachers, preparing training content, and meeting with parents to raise their awareness of cybersecurity topics. All the parents in our study provided positive feedback on these training programs and appreciated the schools' efforts to raise children's cybersecurity awareness.  

\subsection{RQ3. What challenges do parents face in ensuring cybersecurity at home?}
\begin{figure}[h!]
\includegraphics[scale=0.7]{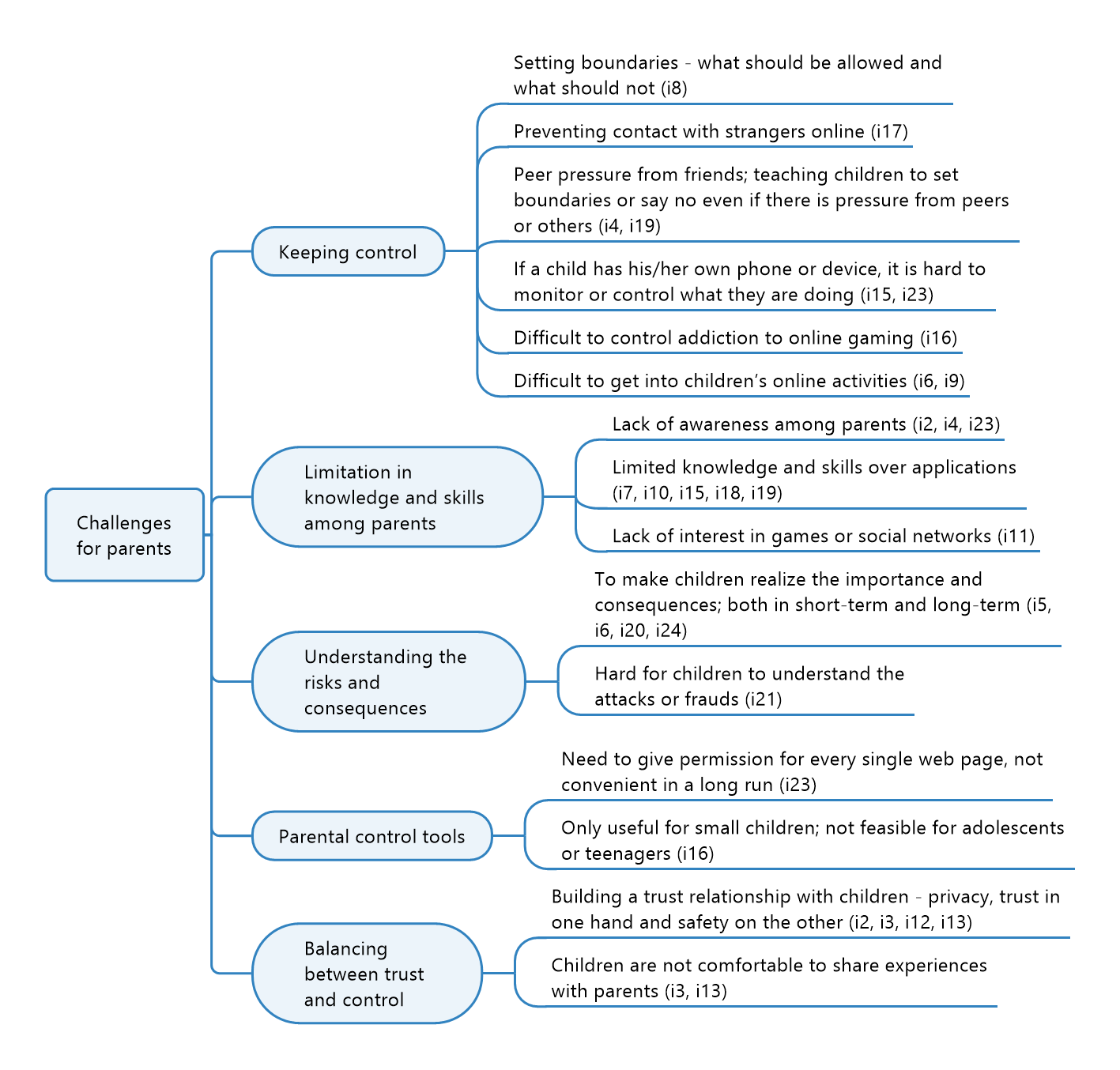}
\caption{Challenges mentioned by the parents in ensuring cybersecurity at home}
\Description{A list of challenges mentioned by the parents}
\label{fig:challenges}
\end{figure}

Throughout the interviews, all the parents shared their feelings about the challenges they faced in ensuring their children's security online. We categorized these challenges based on their focus: keeping control, limitation of knowledge, understanding the risks, limitation of parental control tools, and balancing between trust and control (see Fig. \ref{fig:challenges}).

Overall, all the parents in our study mentioned facing challenges in ensuring their children's security online. Nine of the 25 parents mentioned that they find it difficult to control their children's online activities (e.g., contacting a stranger). Parents mentioned some reasons why they find it difficult to control their children's activity. These reasons include their children's use of personal mobile devices, peer pressure, and the easy and open nature of software such as online games that allow users to contact others without restriction. The second-most common challenge is the parents' lack of knowledge and skills. Nine of the 25 parents mentioned that they lack the knowledge and skills necessary to protect their children from online risks.

The majority of the parents (18/25 parents) in our study have mentioned using parental control, mostly the default parental control features provided by the devices. However, two of the parents have expressed having challenges related to the parental control systems. These challenges are related to the system's feasibility for older children (the teenagers) and their functionality. The parents use parental controls mainly to limit screen time, limit access to different websites, and limit download opportunities.

\section{Discussion}
\label{discussion}
In this section, we discuss the main findings from our research in relation to the research questions of this study. 

\subsection{Cybersecurity awareness needs for children}
As gatekeepers of children's use of technology and the internet, all the parents we interviewed were concerned about their children's safety online. Parents are aware that children can still be at risk without proper awareness, despite the parents' strategies at home. Parents discussed their concerns about cybersecurity risks such as privacy, cyberbullying, stranger danger (e.g., impersonation and grooming), online content, and technical threats (e.g., hacking, malware, and viruses).  

However, it is important to note that some parents expressed some awareness needs for their children that had not been mentioned or studied in previous studies. For example, three parents said that children should understand the technologies they use. Children should understand how the system works, its functionalities, how information disseminates through the system, what information the system collects from users, and how the data is stored. Parents believe that understanding the system will help children to use the technologies in a more secure and mature way. We recommend that the designers of software for children consider these needs. Presenting the privacy policy in a child-friendly way, simplifying user guidelines, and providing instructional notes either inside the system or as a separate user manual can help children to gain a clearer understanding of the system. 

Another interesting finding involves specific concerns about online content. While previous studies have focused mainly on issues like pornography \cite{p25, p34, p49}, age-inappropriate content \cite{p80, p5, p8}, and spam \cite{p33, p70}, some parents (7 of the 25 parents) in our study mentioned that children also need to be aware of the reliability and credibility of online content, intellectual property laws, and the legality of online content. Some parents shared their concerns about how online content can negatively influence children.

Additionally, some parents (3 of the 25) also described the need for children to be positively skeptical when something unusual happens online. This skepticism can also help them to think critically and pause before pressing "OK." With positive skepticism, children can also limit the online sources they accept. However, positive skepticism cannot be directly taught in schools or other organizations. Rather, parents or adults can help children to understand what positive skepticism is and how it is a key aspect of critical thinking. It is also important to help children to differentiate between positive and negative skepticism. Discussing critical thinking and positive skepticism with children can be a good strategy, just like discussing cybersecurity risks and consequences.  

\subsection{Children's knowledge of cybersecurity and their sources of learning}
Aligned with the literature \cite{p5, p65, p21}, in our study, we have seen both parents and teachers playing an essential role in increasing cybersecurity awareness among children in Norway. Most parents in our study believed that their children who are around 10 years old or above have some awareness of cybersecurity. This belief can also be supported by findings from other studies \cite{p8, p36, p55}. Parents also mentioned that there is a difference between knowing about cybersecurity and actually being safe online. Moreover, there is probably much that children do not know about cybersecurity. Thus, they still require more training on this topic. At home, parents try to increase their children's awareness mainly by discussing the relevant security topics with them. Parents often use cybersecurity-related cases from the media as examples in their discussions. 

Here, it is worth noting the initiatives of schools. From the interviews with parents, it became evident that Norwegian schools have taken major steps towards increasing children's cybersecurity awareness. The schools collaborate with organizations that have expertise in the area, aiming to increase cybersecurity awareness not only among children but also among teachers and parents. Parents mentioned many cybersecurity topics that their children learned in school, including safe password practices, privacy, information sharing, contacting strangers, sexual abuse, social media, and cyberbullying. The schools arrange meetings and seminars where parents can learn about cybersecurity and share their experiences. We believe such initiatives can be even more effective if carried out in collaboration with government regulatory bodies and other stakeholders, such as the tech industry. These Norwegian initiatives can serve as an example for other countries seeking to develop frameworks and initiatives to raise cybersecurity awareness among children.

\subsection{Challenges for parents}
Previous studies have demonstrated that parents can play an essential role in shaping children's online behavior and in making them aware of cybersecurity \cite{p65, p73, sonia}. Thus, it is necessary that parents be adequately trained on the topic. If parents receive relevant training and knowledge, they will better understand cybersecurity risks and consequences, and they will be able to help their children with greater confidence. Researchers and other stakeholders, such as the tech industry, can provide parents with the necessary training and guidelines about what to do and what not to do as parents in addressing the challenges. Although Norwegian parents, like parents in other Scandinavian countries, are less authoritative and prefer modern child-rearing attitudes, as shown in \cite{marc}, they still face challenges in balancing between trust and control when monitoring their children's online activities. In terms of building trusting relationships with children and balancing between trust and control, social and behavioral scientists can play a significant role by exploring the parent-child relationship when addressing cybersecurity in the home.

Over the past few years, using parental controls has becoming increasingly popular. Most devices now have some form of parental control built in, enabling parents to regulate their children's data use and access to certain content. \cite{wisniewski3} conducted a feature analysis of 75 Android mobile apps designed to promote adolescent online safety. They found that parental control apps strongly favored features that promoted parental control through monitoring and restricting internet access, rather than encouraging self-regulation or cooperation between parents and teenagers. We can connect this finding to the challenges mentioned by the parents from our study regarding the feasibility and functionality of parental control apps. Theoretically, parental control apps can be useful for children up to age 18. In practice, however, it is difficult for parents to control or restrict children's online activities once they become teenagers. To solve the challenges reported by the parents, we suggest involving both parents and children in the design process when developing parental control apps. Involving children in the design process along with parents can help designers to understand the perspectives of both user groups; the apps will thus better reflect the expectations of both parents and children. Involving parents and children will also encourage collaboration and communication between them, resulting in an increased understanding and a more effective use of parental controls.

\subsection{Limitation and future work}
This study's sample primarily included highly educated parents. Although we did not set any criteria concerning the parents' educational qualifications, after the interviews, it appeared that all the parents in our interviews were highly educated. This self-selection bias that accompanies interview-based research may have influenced the results of this study. Highly educated parents may be more concerned about cybersecurity than less-educated parents. Future research can study the perceptions of parents from more diverse educational backgrounds and identify significant differences from the results of this study. 

In our future work, we aim to develop a framework to equip children and their parents with necessary cybersecurity knowledge and skills. This study has helped us to understand Norwegian parents' views and experiences of cybersecurity awareness for children. This knowledge will help us to develop a cybersecurity awareness framework for children that incorporates their parents' views and needs. 

\section{Conclusion}
\label{conclusion}
In this paper, we presented the results of a qualitative study to understand parents' perspectives of cybersecurity and the cybersecurity awareness needs of children. Parents in our study believe that children need to understand the technologies they use, including privacy issues, online content, online contacts, and also the long-term and short-term consequences of their actions. Children also need some behavioral attributes, such positive skepticism and proper "netiquette," to stay safe online. We recommend that future research consider parents' opinions and concerns when developing tools and products to raise cybersecurity awareness among children. 

We found that children in Norway have some awareness of cybersecurity depending on their age, and they get this awareness from their families and from schools. Parents discuss cybersecurity topics at home frequently or occasionally, whereas schools arrange training programs in cooperation with other organizations. We also explored the challenges for parents in ensuring their children's security online. Despite using different strategies, parents face various challenges in protecting their children from cybersecurity risks. We suggest that parents also need awareness, training, and guidelines on cybersecurity. Parents may benefit from such training on how to help children to develop good privacy and security practices. This can prepare children to better manage their cybersecurity and privacy needs when they begin using the internet and internet-connected devices. 

\begin{acks}
We thank all the parents who participated in the interviews and made our research possible.  
\end{acks}

\bibliographystyle{ACM-Reference-Format}
\bibliography{references}


\begin{thebibliography}{29}


\ifx \showCODEN    \undefined \def \showCODEN     #1{\unskip}     \fi
\ifx \showDOI      \undefined \def \showDOI       #1{#1}\fi
\ifx \showISBNx    \undefined \def \showISBNx     #1{\unskip}     \fi
\ifx \showISBNxiii \undefined \def \showISBNxiii  #1{\unskip}     \fi
\ifx \showISSN     \undefined \def \showISSN      #1{\unskip}     \fi
\ifx \showLCCN     \undefined \def \showLCCN      #1{\unskip}     \fi
\ifx \shownote     \undefined \def \shownote      #1{#1}          \fi
\ifx \showarticletitle \undefined \def \showarticletitle #1{#1}   \fi
\ifx \showURL      \undefined \def \showURL       {\relax}        \fi
\providecommand\bibfield[2]{#2}
\providecommand\bibinfo[2]{#2}
\providecommand\natexlab[1]{#1}
\providecommand\showeprint[2][]{arXiv:#2}

\bibitem[\protect\citeauthoryear{Abd~Rahim, Hamid, Kiah, Shamshirband, and
  Furnell}{Abd~Rahim et~al\mbox{.}}{2015}]%
        {rahim}
\bibfield{author}{\bibinfo{person}{Noor~Hayani Abd~Rahim},
  \bibinfo{person}{Suraya Hamid}, \bibinfo{person}{Miss Laiha~Mat Kiah},
  \bibinfo{person}{Shahaboddin Shamshirband}, {and} \bibinfo{person}{Steven
  Furnell}.} \bibinfo{year}{2015}\natexlab{}.
\newblock \showarticletitle{A systematic review of approaches to assessing
  cybersecurity awareness}.
\newblock \bibinfo{journal}{\emph{Kybernetes}} (\bibinfo{year}{2015}).
\newblock


\bibitem[\protect\citeauthoryear{Ahmad, Arifin, Mokhtar, Hood, Tiun, and
  Jambari}{Ahmad et~al\mbox{.}}{2019}]%
        {p25}
\bibfield{author}{\bibinfo{person}{Nazilah Ahmad}, \bibinfo{person}{Ahmad
  Arifin}, \bibinfo{person}{Umi~Asma Mokhtar}, \bibinfo{person}{Zaihosnita
  Hood}, \bibinfo{person}{Sabrina Tiun}, {and} \bibinfo{person}{Dian~Indrayani
  Jambari}.} \bibinfo{year}{2019}\natexlab{}.
\newblock \showarticletitle{Parental Awareness on Cyber Threats Using Social
  Media}.
\newblock \bibinfo{journal}{\emph{Jurnal Komunikasi: Malaysian Journal of
  Communication}} \bibinfo{volume}{35}, \bibinfo{number}{2}
  (\bibinfo{year}{2019}), \bibinfo{pages}{485--498}.
\newblock
\urldef\tempurl%
\url{https://doi.org/10.17576/JKMJC-2019-3502-29}
\showDOI{\tempurl}


\bibitem[\protect\citeauthoryear{Baciu-Ureche, Sleeman, Moody, and
  Matthews}{Baciu-Ureche et~al\mbox{.}}{2019}]%
        {p24}
\bibfield{author}{\bibinfo{person}{Ovidiu-Gabriel Baciu-Ureche},
  \bibinfo{person}{Carlie Sleeman}, \bibinfo{person}{William~C. Moody}, {and}
  \bibinfo{person}{Suzanne~J. Matthews}.} \bibinfo{year}{2019}\natexlab{}.
\newblock \showarticletitle{The Adventures of ScriptKitty: Using the Raspberry
  Pi to Teach Adolescents about Internet Safety}. In
  \bibinfo{booktitle}{\emph{Proceedings of the 20th Annual SIG Conference on
  Information Technology Education}} \emph{(\bibinfo{series}{SIGITE '19})}.
  \bibinfo{pages}{118–123}.
\newblock
\showISBNx{9781450369213}
\urldef\tempurl%
\url{https://doi.org/10.1145/3349266.3351399}
\showDOI{\tempurl}


\bibitem[\protect\citeauthoryear{Basili, Caldiera, and Rombach}{Basili
  et~al\mbox{.}}{1994}]%
        {basili}
\bibfield{author}{\bibinfo{person}{Victor~R. Basili},
  \bibinfo{person}{Gianluigi Caldiera}, {and} \bibinfo{person}{H.~Dieter
  Rombach}.} \bibinfo{year}{1994}\natexlab{}.
\newblock \showarticletitle{THE GOAL QUESTION METRIC APPROACH}.
\newblock \bibinfo{journal}{\emph{Encyclopedia of Software Engineering}}
  (\bibinfo{year}{1994}).
\newblock


\bibitem[\protect\citeauthoryear{{Bioglio}, {Capecchi}, {Peiretti}, {Sayed},
  {Torasso}, and {Pensa}}{{Bioglio} et~al\mbox{.}}{2019}]%
        {p21}
\bibfield{author}{\bibinfo{person}{L. {Bioglio}}, \bibinfo{person}{S.
  {Capecchi}}, \bibinfo{person}{F. {Peiretti}}, \bibinfo{person}{D. {Sayed}},
  \bibinfo{person}{A. {Torasso}}, {and} \bibinfo{person}{R.~G. {Pensa}}.}
  \bibinfo{year}{2019}\natexlab{}.
\newblock \showarticletitle{A Social Network Simulation Game to Raise Awareness
  of Privacy Among School Children}.
\newblock \bibinfo{journal}{\emph{IEEE Transactions on Learning Technologies}}
  \bibinfo{volume}{12}, \bibinfo{number}{4} (\bibinfo{year}{2019}),
  \bibinfo{pages}{456--469}.
\newblock
\urldef\tempurl%
\url{https://doi.org/10.1109/TLT.2018.2881193}
\showDOI{\tempurl}


\bibitem[\protect\citeauthoryear{Bornstein, Putnick, and Lansford}{Bornstein
  et~al\mbox{.}}{2011}]%
        {marc}
\bibfield{author}{\bibinfo{person}{Marc~H. Bornstein},
  \bibinfo{person}{Diane~L. Putnick}, {and} \bibinfo{person}{Jennifer~E.
  Lansford}.} \bibinfo{year}{2011}\natexlab{}.
\newblock \showarticletitle{Parenting Attributions and Attitudes in
  Cross-Cultural Perspective}.
\newblock \bibinfo{journal}{\emph{Parenting}} \bibinfo{volume}{11},
  \bibinfo{number}{2-3} (\bibinfo{year}{2011}), \bibinfo{pages}{214--237}.
\newblock
\urldef\tempurl%
\url{https://doi.org/10.1080/15295192.2011.585568}
\showDOI{\tempurl}


\bibitem[\protect\citeauthoryear{Clausen}{Clausen}{1996}]%
        {clausen1996parenting}
\bibfield{author}{\bibinfo{person}{Sten-Erik Clausen}.}
  \bibinfo{year}{1996}\natexlab{}.
\newblock \showarticletitle{Parenting styles and adolescent drug use
  behaviours}.
\newblock \bibinfo{journal}{\emph{Childhood}} \bibinfo{volume}{3},
  \bibinfo{number}{3} (\bibinfo{year}{1996}), \bibinfo{pages}{403--414}.
\newblock


\bibitem[\protect\citeauthoryear{Clemons and Wilson}{Clemons and
  Wilson}{2015}]%
        {p84}
\bibfield{author}{\bibinfo{person}{Eric~K. Clemons} {and}
  \bibinfo{person}{Joshua~S. Wilson}.} \bibinfo{year}{2015}\natexlab{}.
\newblock \showarticletitle{Family Preferences Concerning Online Privacy, Data
  Mining, and Targeted Ads: Regulatory Implications}.
\newblock \bibinfo{journal}{\emph{Journal of Management Information Systems}}
  \bibinfo{volume}{32}, \bibinfo{number}{2} (\bibinfo{year}{2015}),
  \bibinfo{pages}{40--70}.
\newblock
\urldef\tempurl%
\url{https://doi.org/10.1080/07421222.2015.1063277}
\showDOI{\tempurl}


\bibitem[\protect\citeauthoryear{{Cruzes} and {Dyb\aa}}{{Cruzes} and
  {Dyb\aa}}{2011}]%
        {cruzes-dyba}
\bibfield{author}{\bibinfo{person}{Daniela.~S. {Cruzes}} {and}
  \bibinfo{person}{Tore {Dyb\aa}}.} \bibinfo{year}{2011}\natexlab{}.
\newblock \showarticletitle{Recommended Steps for Thematic Synthesis in
  Software Engineering}. In \bibinfo{booktitle}{\emph{2011 International
  Symposium on Empirical Software Engineering and Measurement}}.
  \bibinfo{pages}{275--284}.
\newblock


\bibitem[\protect\citeauthoryear{Elstad and Stefansen}{Elstad and
  Stefansen}{2014}]%
        {elstad2014social}
\bibfield{author}{\bibinfo{person}{Jon~Ivar Elstad} {and} \bibinfo{person}{Kari
  Stefansen}.} \bibinfo{year}{2014}\natexlab{}.
\newblock \showarticletitle{Social variations in perceived parenting styles
  among Norwegian adolescents}.
\newblock \bibinfo{journal}{\emph{Child indicators research}}
  \bibinfo{volume}{7}, \bibinfo{number}{3} (\bibinfo{year}{2014}),
  \bibinfo{pages}{649--670}.
\newblock


\bibitem[\protect\citeauthoryear{Giannakas, Kambourakis, Papasalouros, and
  Gritzalis}{Giannakas et~al\mbox{.}}{2016}]%
        {p70}
\bibfield{author}{\bibinfo{person}{Filippos Giannakas},
  \bibinfo{person}{Georgios Kambourakis}, \bibinfo{person}{Andreas
  Papasalouros}, {and} \bibinfo{person}{Stefanos Gritzalis}.}
  \bibinfo{year}{2016}\natexlab{}.
\newblock \showarticletitle{Security Education and Awareness for K-6 Going
  Mobile}.
\newblock \bibinfo{journal}{\emph{International Journal of Interactive Mobile
  Technologies (iJIM)}} \bibinfo{volume}{10}, \bibinfo{number}{2}
  (\bibinfo{year}{2016}), \bibinfo{pages}{41--48}.
\newblock
\showISSN{1865-7923}


\bibitem[\protect\citeauthoryear{{Kritzinger}}{{Kritzinger}}{2015}]%
        {p80}
\bibfield{author}{\bibinfo{person}{E. {Kritzinger}}.}
  \bibinfo{year}{2015}\natexlab{}.
\newblock \showarticletitle{Enhancing cyber safety awareness among school
  children in South Africa through gaming}. In \bibinfo{booktitle}{\emph{2015
  Science and Information Conference (SAI)}}. \bibinfo{pages}{1243--1248}.
\newblock
\urldef\tempurl%
\url{https://doi.org/10.1109/SAI.2015.7237303}
\showDOI{\tempurl}


\bibitem[\protect\citeauthoryear{Kumar, Naik, Devkar, Chetty, Clegg, and
  Vitak}{Kumar et~al\mbox{.}}{2017}]%
        {p55}
\bibfield{author}{\bibinfo{person}{Priya Kumar},
  \bibinfo{person}{Shalmali~Milind Naik}, \bibinfo{person}{Utkarsha~Ramesh
  Devkar}, \bibinfo{person}{Marshini Chetty}, \bibinfo{person}{Tamara~L.
  Clegg}, {and} \bibinfo{person}{Jessica Vitak}.}
  \bibinfo{year}{2017}\natexlab{}.
\newblock \showarticletitle{No Telling Passcodes Out Because They're Private:
  Understanding Children's Mental Models of Privacy and Security Online}.
\newblock \bibinfo{journal}{\emph{Proc. ACM Hum.-Comput. Interact.}}
  \bibinfo{volume}{1}, \bibinfo{number}{CSCW}, Article \bibinfo{articleno}{64}
  (\bibinfo{date}{Dec.} \bibinfo{year}{2017}), \bibinfo{numpages}{21}~pages.
\newblock
\urldef\tempurl%
\url{https://doi.org/10.1145/3134699}
\showDOI{\tempurl}


\bibitem[\protect\citeauthoryear{Kumar, Vitak, Chetty, Clegg, Yang, McNally,
  and Bonsignore}{Kumar et~al\mbox{.}}{2018}]%
        {p36}
\bibfield{author}{\bibinfo{person}{Priya Kumar}, \bibinfo{person}{Jessica
  Vitak}, \bibinfo{person}{Marshini Chetty}, \bibinfo{person}{Tamara~L. Clegg},
  \bibinfo{person}{Jonathan Yang}, \bibinfo{person}{Brenna McNally}, {and}
  \bibinfo{person}{Elizabeth Bonsignore}.} \bibinfo{year}{2018}\natexlab{}.
\newblock \showarticletitle{Co-Designing Online Privacy-Related Games and
  Stories with Children}. In \bibinfo{booktitle}{\emph{Proceedings of the 17th
  ACM Conference on Interaction Design and Children}} (Trondheim, Norway)
  \emph{(\bibinfo{series}{IDC '18})}. \bibinfo{pages}{67–79}.
\newblock
\showISBNx{9781450351522}
\urldef\tempurl%
\url{https://doi.org/10.1145/3202185.3202735}
\showDOI{\tempurl}


\bibitem[\protect\citeauthoryear{Lastdrager, Gallardo, Hartel, and
  Junger}{Lastdrager et~al\mbox{.}}{2017}]%
        {p14}
\bibfield{author}{\bibinfo{person}{Elmer Lastdrager},
  \bibinfo{person}{́Ines~Carvajal Gallardo}, \bibinfo{person}{Pieter Hartel},
  {and} \bibinfo{person}{Marianne Junger}.} \bibinfo{year}{2017}\natexlab{}.
\newblock \showarticletitle{How Effective is Anti-Phishing Training for
  Children?}. In \bibinfo{booktitle}{\emph{Thirteenth Symposium on Usable
  Privacy and Security ({SOUPS} 2017)}}. \bibinfo{publisher}{{USENIX}
  Association}, \bibinfo{address}{Santa Clara, CA}, \bibinfo{pages}{229--239}.
\newblock
\showISBNx{978-1-931971-39-3}


\bibitem[\protect\citeauthoryear{Livingstone and Helsper}{Livingstone and
  Helsper}{2008}]%
        {sonia}
\bibfield{author}{\bibinfo{person}{Sonia Livingstone} {and}
  \bibinfo{person}{Ellen~J. Helsper}.} \bibinfo{year}{2008}\natexlab{}.
\newblock \showarticletitle{Parental Mediation of Children's Internet Use}.
\newblock \bibinfo{journal}{\emph{Journal of Broadcasting \& Electronic Media}}
  \bibinfo{volume}{52}, \bibinfo{number}{4} (\bibinfo{year}{2008}),
  \bibinfo{pages}{581--599}.
\newblock
\urldef\tempurl%
\url{https://doi.org/10.1080/08838150802437396}
\showDOI{\tempurl}
\showeprint{https://doi.org/10.1080/08838150802437396}


\bibitem[\protect\citeauthoryear{Maoneke, Shava, Gamundani, Bere-Chitauro, and
  Nhamu}{Maoneke et~al\mbox{.}}{2018}]%
        {p34}
\bibfield{author}{\bibinfo{person}{Pardon~Blessings Maoneke},
  \bibinfo{person}{Fungai~Bhunu Shava}, \bibinfo{person}{Attlee~Munyaradzi
  Gamundani}, \bibinfo{person}{Mercy Bere-Chitauro}, {and}
  \bibinfo{person}{Isaac Nhamu}.} \bibinfo{year}{2018}\natexlab{}.
\newblock \showarticletitle{ICTs Use and Cyberspace Risks Faced by Adolescents
  in Namibia}. In \bibinfo{booktitle}{\emph{Proceedings of the Second African
  Conference for Human Computer Interaction: Thriving Communities}} (Windhoek,
  Namibia) \emph{(\bibinfo{series}{AfriCHI '18})}. Article
  \bibinfo{articleno}{11}, \bibinfo{numpages}{9}~pages.
\newblock
\showISBNx{9781450365581}
\urldef\tempurl%
\url{https://doi.org/10.1145/3283458.3283483}
\showDOI{\tempurl}


\bibitem[\protect\citeauthoryear{Martin, Wang, Petty, Wang, and Wilkins}{Martin
  et~al\mbox{.}}{2018}]%
        {p33}
\bibfield{author}{\bibinfo{person}{Florence Martin}, \bibinfo{person}{Chuang
  Wang}, \bibinfo{person}{Teresa Petty}, \bibinfo{person}{Weichao Wang}, {and}
  \bibinfo{person}{Patti Wilkins}.} \bibinfo{year}{2018}\natexlab{}.
\newblock \showarticletitle{Middle School Students’ Social Media Use}.
\newblock \bibinfo{journal}{\emph{Journal of Educational Technology and
  Society}} \bibinfo{volume}{21}, \bibinfo{number}{1} (\bibinfo{year}{2018}),
  \bibinfo{pages}{213--224}.
\newblock
\showISSN{11763647, 14364522}
\urldef\tempurl%
\url{http://www.jstor.org/stable/26273881}
\showURL{%
\tempurl}


\bibitem[\protect\citeauthoryear{Moreno, Egan, and Bare}{Moreno
  et~al\mbox{.}}{2013}]%
        {p98}
\bibfield{author}{\bibinfo{person}{M.A. Moreno}, \bibinfo{person}{K.G. Egan},
  {and} \bibinfo{person}{K Bare}.} \bibinfo{year}{2013}\natexlab{}.
\newblock \showarticletitle{Internet safety education for youth: stakeholder
  perspectives}.
\newblock \bibinfo{journal}{\emph{BMC Public Health}} \bibinfo{volume}{13},
  \bibinfo{number}{543} (\bibinfo{year}{2013}).
\newblock
\showISSN{1471-2458}
\urldef\tempurl%
\url{https://doi.org/10.1186/1471-2458-13-543}
\showDOI{\tempurl}


\bibitem[\protect\citeauthoryear{Muir and Joinson}{Muir and Joinson}{2020}]%
        {p5}
\bibfield{author}{\bibinfo{person}{Kate Muir} {and} \bibinfo{person}{Adam
  Joinson}.} \bibinfo{year}{2020}\natexlab{}.
\newblock \showarticletitle{An Exploratory Study Into the Negotiation of
  Cyber-Security Within the Family Home}.
\newblock \bibinfo{journal}{\emph{Frontiers in Psychology}}
  \bibinfo{volume}{11} (\bibinfo{year}{2020}), \bibinfo{pages}{424}.
\newblock
\showISSN{1664-1078}
\urldef\tempurl%
\url{https://doi.org/10.3389/fpsyg.2020.00424}
\showDOI{\tempurl}


\bibitem[\protect\citeauthoryear{Prior and Renaud}{Prior and Renaud}{2020}]%
        {p6}
\bibfield{author}{\bibinfo{person}{Suzanne Prior} {and} \bibinfo{person}{Karen
  Renaud}.} \bibinfo{year}{2020}\natexlab{}.
\newblock \showarticletitle{Age-appropriate password “best practice”
  ontologies for early educators and parents}.
\newblock \bibinfo{journal}{\emph{International Journal of Child-Computer
  Interaction}}  \bibinfo{volume}{23-24} (\bibinfo{year}{2020}),
  \bibinfo{pages}{100169}.
\newblock
\showISSN{2212-8689}
\urldef\tempurl%
\url{https://doi.org/10.1016/j.ijcci.2020.100169}
\showDOI{\tempurl}


\bibitem[\protect\citeauthoryear{Rode}{Rode}{2009}]%
        {rode}
\bibfield{author}{\bibinfo{person}{Jennifer~A. Rode}.}
  \bibinfo{year}{2009}\natexlab{}.
\newblock \showarticletitle{Digital Parenting: Designing Children's Safety}. In
  \bibinfo{booktitle}{\emph{Proceedings of the 23rd British HCI Group Annual
  Conference on People and Computers: Celebrating People and Technology}}
  (Cambridge, United Kingdom) \emph{(\bibinfo{series}{BCS-HCI '09})}.
  \bibinfo{publisher}{BCS Learning and Development Ltd.},
  \bibinfo{address}{Swindon, GBR}, \bibinfo{pages}{244–251}.
\newblock


\bibitem[\protect\citeauthoryear{Shin and Kang}{Shin and Kang}{2016}]%
        {p65}
\bibfield{author}{\bibinfo{person}{Wonsun Shin} {and} \bibinfo{person}{Hyunjin
  Kang}.} \bibinfo{year}{2016}\natexlab{}.
\newblock \showarticletitle{Adolescents' privacy concerns and information
  disclosure online: The role of parents and the Internet}.
\newblock \bibinfo{journal}{\emph{Computers in Human Behavior}}
  \bibinfo{volume}{54} (\bibinfo{year}{2016}), \bibinfo{pages}{114 -- 123}.
\newblock
\showISSN{0747-5632}
\urldef\tempurl%
\url{https://doi.org/10.1016/j.chb.2015.07.062}
\showDOI{\tempurl}


\bibitem[\protect\citeauthoryear{Valcke, Bonte, {De Wever}, and Rots}{Valcke
  et~al\mbox{.}}{2010}]%
        {VALCKE2010454}
\bibfield{author}{\bibinfo{person}{M. Valcke}, \bibinfo{person}{S. Bonte},
  \bibinfo{person}{B. {De Wever}}, {and} \bibinfo{person}{I. Rots}.}
  \bibinfo{year}{2010}\natexlab{}.
\newblock \showarticletitle{Internet parenting styles and the impact on
  Internet use of primary school children}.
\newblock \bibinfo{journal}{\emph{Computers \& Education}}
  \bibinfo{volume}{55}, \bibinfo{number}{2} (\bibinfo{year}{2010}),
  \bibinfo{pages}{454--464}.
\newblock
\showISSN{0360-1315}
\urldef\tempurl%
\url{https://doi.org/10.1016/j.compedu.2010.02.009}
\showDOI{\tempurl}


\bibitem[\protect\citeauthoryear{Wisniewski, Ghosh, Xu, Rosson, and
  Carroll}{Wisniewski et~al\mbox{.}}{2017a}]%
        {wisniewski3}
\bibfield{author}{\bibinfo{person}{Pamela Wisniewski},
  \bibinfo{person}{Arup~Kumar Ghosh}, \bibinfo{person}{Heng Xu},
  \bibinfo{person}{Mary~Beth Rosson}, {and} \bibinfo{person}{John~M. Carroll}.}
  \bibinfo{year}{2017}\natexlab{a}.
\newblock \showarticletitle{Parental Control vs. Teen Self-Regulation: Is There
  a Middle Ground for Mobile Online Safety?}. In
  \bibinfo{booktitle}{\emph{Proceedings of the 2017 ACM Conference on Computer
  Supported Cooperative Work and Social Computing}} (Portland, Oregon, USA)
  \emph{(\bibinfo{series}{CSCW '17})}. \bibinfo{pages}{51–69}.
\newblock
\showISBNx{9781450343350}
\urldef\tempurl%
\url{https://doi.org/10.1145/2998181.2998352}
\showDOI{\tempurl}


\bibitem[\protect\citeauthoryear{Wisniewski, Jia, Xu, Rosson, and
  Carroll}{Wisniewski et~al\mbox{.}}{2015}]%
        {p73}
\bibfield{author}{\bibinfo{person}{Pamela Wisniewski}, \bibinfo{person}{Haiyan
  Jia}, \bibinfo{person}{Heng Xu}, \bibinfo{person}{Mary~Beth Rosson}, {and}
  \bibinfo{person}{John~M. Carroll}.} \bibinfo{year}{2015}\natexlab{}.
\newblock \showarticletitle{"Preventative" vs. "Reactive": How Parental
  Mediation Influences Teens' Social Media Privacy Behaviors}. In
  \bibinfo{booktitle}{\emph{Proceedings of the 18th ACM Conference on Computer
  Supported Cooperative Work and Social Computing}} (Vancouver, BC, Canada)
  \emph{(\bibinfo{series}{CSCW '15})}. \bibinfo{pages}{302–316}.
\newblock
\showISBNx{9781450329224}
\urldef\tempurl%
\url{https://doi.org/10.1145/2675133.2675293}
\showDOI{\tempurl}


\bibitem[\protect\citeauthoryear{Wisniewski, Xu, Rosson, and
  Carroll}{Wisniewski et~al\mbox{.}}{2017b}]%
        {p49}
\bibfield{author}{\bibinfo{person}{Pamela Wisniewski}, \bibinfo{person}{Heng
  Xu}, \bibinfo{person}{Mary~Beth Rosson}, {and} \bibinfo{person}{John~M.
  Carroll}.} \bibinfo{year}{2017}\natexlab{b}.
\newblock \showarticletitle{Parents Just Don't Understand: Why Teens Don't Talk
  to Parents about Their Online Risk Experiences}. In
  \bibinfo{booktitle}{\emph{Proceedings of the 2017 ACM Conference on Computer
  Supported Cooperative Work and Social Computing}} (Portland, Oregon, USA)
  \emph{(\bibinfo{series}{CSCW '17})}. \bibinfo{pages}{523–540}.
\newblock
\showISBNx{9781450343350}
\urldef\tempurl%
\url{https://doi.org/10.1145/2998181.2998236}
\showDOI{\tempurl}


\bibitem[\protect\citeauthoryear{Zhang-Kennedy, Abdelaziz, and
  Chiasson}{Zhang-Kennedy et~al\mbox{.}}{2017}]%
        {p48}
\bibfield{author}{\bibinfo{person}{Leah Zhang-Kennedy}, \bibinfo{person}{Yomna
  Abdelaziz}, {and} \bibinfo{person}{Sonia Chiasson}.}
  \bibinfo{year}{2017}\natexlab{}.
\newblock \showarticletitle{Cyberheroes: The design and evaluation of an
  interactive ebook to educate children about online privacy}.
\newblock \bibinfo{journal}{\emph{International Journal of Child-Computer
  Interaction}}  \bibinfo{volume}{13} (\bibinfo{year}{2017}),
  \bibinfo{pages}{10 -- 18}.
\newblock
\showISSN{2212-8689}
\urldef\tempurl%
\url{https://doi.org/10.1016/j.ijcci.2017.05.001}
\showDOI{\tempurl}


\bibitem[\protect\citeauthoryear{Zhao, Wang, Dally, Slovak, Edbrooke-Childs,
  Van~Kleek, and Shadbolt}{Zhao et~al\mbox{.}}{2019}]%
        {p8}
\bibfield{author}{\bibinfo{person}{Jun Zhao}, \bibinfo{person}{Ge Wang},
  \bibinfo{person}{Carys Dally}, \bibinfo{person}{Petr Slovak},
  \bibinfo{person}{Julian Edbrooke-Childs}, \bibinfo{person}{Max Van~Kleek},
  {and} \bibinfo{person}{Nigel Shadbolt}.} \bibinfo{year}{2019}\natexlab{}.
\newblock \showarticletitle{`I Make up a Silly Name': Understanding Children's
  Perception of Privacy Risks Online}. In \bibinfo{booktitle}{\emph{Proceedings
  of the 2019 CHI Conference on Human Factors in Computing Systems}}
  \emph{(\bibinfo{series}{CHI '19})}. \bibinfo{pages}{1–13}.
\newblock
\showISBNx{9781450359702}
\urldef\tempurl%
\url{https://doi.org/10.1145/3290605.3300336}
\showDOI{\tempurl}


\end{thebibliography}

\end{document}